\documentclass[journal=jacsat,manuscript=article]{achemso}
\setkeys{acs}{articletitle=true}
\usepackage[version=3]{mhchem} 

\usepackage{graphicx}
\usepackage{dcolumn}
\usepackage{bm}
\usepackage{xcolor}
\usepackage{siunitx}
\usepackage{amsmath}
\newcommand{\kT}{\ensuremath{k_{\rm B}T}}
\renewcommand{\vec}[1]{\mathbf{#1}}

\title[Parallel Biasing in Flat Histogram Methods]{Exploring the Potential of Parallel Biasing in Flat Histogram Methods}

\author{Shanghui Huang}
 \affiliation{Department of Chemistry and Biochemistry,\\ University of Notre Dame, Notre Dame, IN, 46556, USA}
\author{Michael J. Quevillon}
\affiliation{Department of Chemical and Biomolecular Engineering,\\ University of Notre Dame, Notre Dame, IN, 46556, USA}
\author{Ernesto C. Cort\'{e}s-Morales}
\affiliation{Department of Chemical and Biomolecular Engineering,\\ University of Notre Dame, Notre Dame, IN, 46556, USA}
\author{Jonathan K. Whitmer}%
 \email{jwhitme1@nd.edu}
\affiliation{Department of Chemistry and Biochemistry,\\ University of Notre Dame, Notre Dame, IN, 46556, USA}%
\alsoaffiliation{Department of Chemical and Biomolecular Engineering,\\ University of Notre Dame, Notre Dame, IN, 46556, USA}%

\date{\today}
\begin{document}

\begin{abstract}
Metadynamics, a member of the `flat histogram' class of advanced sampling algorithms, has been widely used in molecular simulations to drive the exploration of states separated by high free energy barriers and promote comprehensive sampling of free energy landscapes defined on collective variables (CVs) which characterize the state of the system. Typically, the methods encounter severe limitations when exploring large numbers of CVs. A recently proposed variant, parallel bias metadynamics (PBMetaD), promises to aid in exploring free energy landscapes along with multiple important collective variables by exchanging the $n$-dimensional free energy landscape required by standard methods for $n$ one-dimensional marginal free energy landscapes. In this study, we systematically examine how parallel biasing affects the convergence of free energy landscapes along with each variable relative to standard methods and the effectiveness of the parallel biasing strategy for addressing common bottlenecks in the use of advanced sampling to calculate free energies.
\end{abstract}

\maketitle

\section{Introduction}
Molecular dynamics (MD) simulation plays a crucial role in scientific studies within many diverse fields, including biology and materials science.\citep{Karplus2002,Yan2004,Zhao2020,Agrawal2020,Wang2020} Robust sampling over the phase space of interest is imperative for obtaining accurate results.  Standard MD simulations, while enormously successful in predicting a wide range of properties and mechanisms, fail to adequately sample configurational space in systems which have even relatively modest ($\mathcal{O}(10\ \kT)$) energy barriers between configurations without running simulations which evolve on a timescale of \SI{}{\micro\second} or longer, which despite ever-improving computer hardware is still computationally expensive.\citep{Klepeis2009,Shaw2010,Lindorff-Larsen2011,Dror2012} This is an acute problem in complex or multiphase media, including protein folding applications or studies of phase transitions in soft materials, where metastable states can trap systems for long periods of time.\citep{Shi2019,Leonhard2019,Shi2020,Rathee2018,Rathee2018a,Quevillon2018,Huang2020} 

One can quantify the stability of specific configurations within a system and the transitions between them using the statistical mechanics of augmented ensembles, informed by collective variables (CVs), which can include the order parameters characterizing phase transitions, or chemical reaction coordinates characterizing the rate limiting steps of a system. The definition of CVs provides a coarse-grained description of the system in reduced dimensions; one may regard the complete phase space as being projected onto this space of CVs, thus lumping conformationally similar states. 

A class of advanced sampling techniques proceeds by applying potential energy biases to these CVs, hoping to facilitate the crossing of free energy barriers. One successful strategy comes from the Metadynamics family of methods (MetaD),\citep{Laio2002,Laio2005,Awasthi2019,Bussi2020} which adds Gaussian-shaped biases to states visited in CV-space on-the-fly, facilitating the escape of free energy minima and the measurement of free energy with little \textit{a priori} knowledge of the free energy landscape. This has been successfully applied in a wide variety of situations; interested readers should consult recent review articles for specific examples.\citep{Bussi2020,Singh2012} In MetaD methods, biasing potentials are updated for visited states within the ensemble until they reach conditions that can be regarded as converged, after which the free energy landscape is mollified and the previously unknown free energy landscape may be determined.

Though free energy calculation is often the goal of a MetaD-enhanced simulation, MetaD can also be used to drive sampling to new physical configurations, which may then be explored in subsequent simulations. In these cases, multiple CVs can be more easily used, since repeated sampling to convergence is not needed. While these aims can be explored simultaneously, the use of MetaD typically involves a binary choice to enhance exploration or directly probe a free energy, since typical implementations of the algorithm prohibit sampling along a large number of CVs simultaneously, as convergence relies on diffusion along the biased surface to return many times to each region of interest in the free energy landscape.

\begin{figure}[t!]
    \centering
    \includegraphics[width=0.45\textwidth]{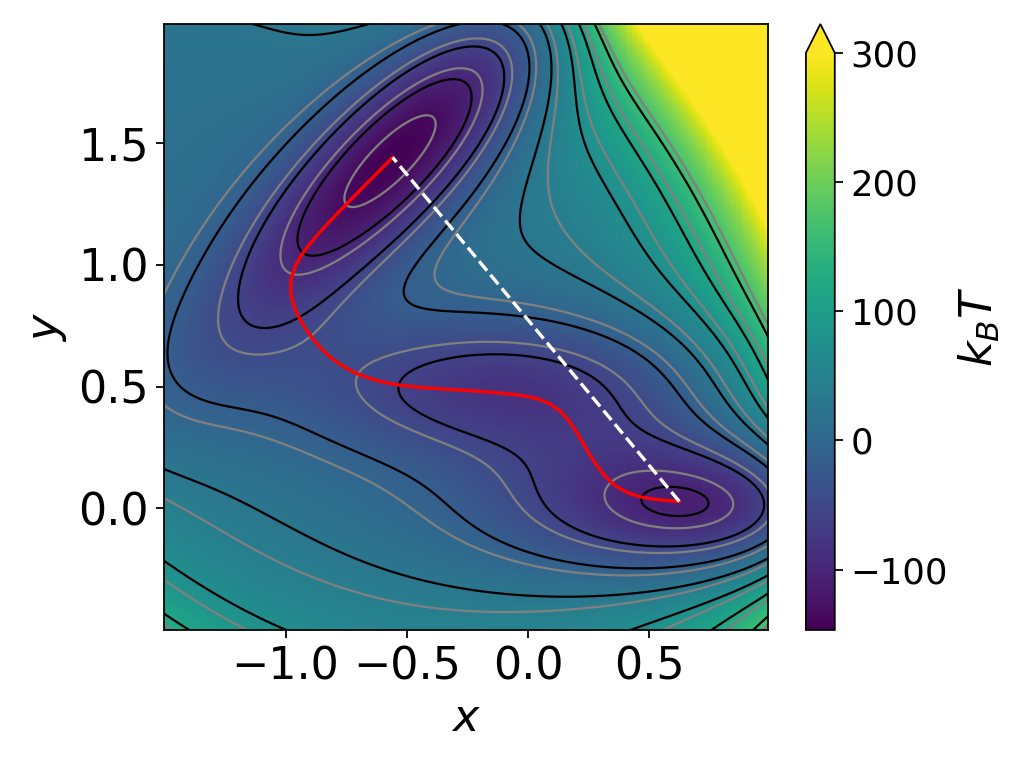}
    \caption{M{\"u}eller potential surface, a structural CV is drawn in white dashed line, the dynamical CV is the red solid curve}
    \label{fig:mueller}
\end{figure}

It should also be noted that without prior knowledge it is difficult to have comprehensive sampling even of systems with only a few CVs.\citep{Chen2018,Chen2018a,Chen2019,Chen2019a} To be more specific, a fundamental requirement of a CV in typical MetaD simulations is that it can distinguish two states of interest; we will denote CVs with this property as {\it structural} CVs. Emphasizing only structural CVs of interest, however, overlooks important information about the dynamics of the system. If {\it structural} CVs do not align with the slowly evolving degrees of freedom or reaction pathways controlling the evolution of a given system, then simple biasing along the structural degrees of freedom may not be enough to achieve swift convergence. Consider the Mueller potential shown in Fig.~\ref{fig:mueller} as an illustrative case.\citep{Muller1980,E2005} Two minima exist in this potential, and a set of structural CVs can be readily identified, including the Cartesian coordinates $x$ and $y$, as well as the straight line connecting the two minima. Each of these projects the two minima into separate locations in CV space, and can be used to identify statistically which minimum the system is spending the most time in. However, the transition path does not lie along any single structural CV; instead, transitions proceed along or near the minimum-energy curve that passes through the saddle point between the two minima. The parametric curve between these sites can be used to identify a CV which faithfully captures the true dynamics of the system. This concept is intimately tied to the concept of reaction coordinates, but they need not be one and the same; dynamical CVs may be used to approximate the reaction coordinate in a piecewise fashion even if the true reaction coordinate is unknown. We will denote such a construction as a {\it dynamical} CV. Note that in a more complex system, multiple {\it dynamical} CVs (or multiple collections of such CVs) may exist, each tying together metastable states in sequence as is necessary for a system to evolve between structural minima. 

Such reaction coordinates are clearly the ideal sampling coordinates for any system, as they combine the two types of CVs into one: reaction coordinates are both structural and dynamic in our classification scheme, as they contain the important dynamical properties in addition to the important thermodynamical ones. However, knowledge of reaction coordinates is extremely elusive, and these typically cannot be known {\it a priori} in order to facilitate sampling.\citep{Chen2018,Chen2018a,Chen2019,Chen2019a,Ribeiro2018,Sultan2018,Mardt2018,Wehmeyer2018} As different CVs can be combined to aid in the dynamics of the system without any of them being the true reaction coordinate, we thus often resort to expanding the set of CVs that are biased on to increase the amount information we can get from a system. Such regulation counters a fact of MetaD that the cost of computational resources increases drastically as the number of CVs increases. A typical examination of a 2D potential like the Mueller potential would note that all information is contained in $x$ and $y$ and so those Cartesian variables can be combined in a multidimensional MetaD simulation to extract all relevant physical information. For simulations with hundreds or thousands of degrees of freedom available, there is often a handful of dynamical CVs which can be combined with structural CVs to balance the need for exploration dynamics with sampling on the structural quantities of interest. This strategy is, however, limited in the total number of degrees of freedom which can be comprehensively explored by the inherent slowness of comprehensive coverage. 

Parallel bias metadynamics (PBMetaD),\citep{Pfaendtner2015} a recently developed variant of Metadynamics, improves sampling efficiency by exchanging $1$ $n$-dimensional for $n$ $1$-dimensional biases, drawing inspiration from the bias-exchange variants of Metadynamics.\citep{Piana2007a} The negative of each bias converges independently to a mollified version of the marginal free energy landscape\bibnote{This comment assumes well-tempering is inherent in the biasing scheme, as is true in our outline of the algorithm within this paper, and for all examples of PBMetaD cited in this paper, which are based on Ref.~\citenum{Pfaendtner2015}.} integrated over all the other degrees of freedom in the system, though the system's exploration is influenced by the current degree of sampling of each individual collective variable. This method enables a user to probe the influence of more CVs than would otherwise be possible in a single MetaD simulation. We hypothesize that this can be utilized to combine driven sampling over degrees of freedom having slow dynamics within the system (thus facilitating transitions) while simultaneously capturing the free energy along a structural coordinate of interest, even when a relatively large number of CVs is necessary to capture all relevant transition states in the system. In this way, the need for the full reaction coordinate is bypassed through PBMetaD sampling of structural and dynamic degrees of freedom which are comparatively simpler to identify (though potentially more numerous). In this work, we apply a systematic study to explore how PBMetaD benefits multiple-CV sampling by enhancing sampling over hidden pathways when {\it structural} CVs are combined with collections of {\it dynamical} CVs.

\section{Method}
For completeness, we briefly recount elements of the PBMetaD method; readers are referred to Ref.~\citenum{Pfaendtner2015} for the complete development. In a conventional MetaD, a small repulsive Gaussian potential is added to the system at intervals of time $\tau_G$ to push the system out of local minima. The bias potential at a time $t$ after the start of the simulation is constructed for a generic multidimensional CV space as 
\begin{equation}
V(\bm{\xi},t) = \sum_{t'\leq t} W(t') e^{-\sum\limits_{\alpha=1}^{d}\frac{[\xi_\alpha-s_{\alpha}(t')]^2}{2\sigma_{\alpha}^2} }   
\end{equation}
\noindent where $\bm{\xi}$ is a $d$-dimensional point in CV space, $s_{\alpha}(t)$ is the instantaneous CV value in the $\alpha$ dimension, and $t'$ is an iterator over times where bias has been applied since the beginning of the simulation which is a multiplier of $\tau_G$ and should be smaller or equal than total simulation time $t$. The function $s(t)$ is a function of the current configuration of the system as determined by the atomic position.  $V(\bm{\xi},t')$ is the total bias added to position $\bm{\xi}$ from the beginning of the simulation until time $t'$, $W(t)$ and $\sigma$ are the time-dependent height and width of the individual Gaussian biases. 

One of the central concerns when MetaD is used to obtain the free energy of a molecular system is determining when the bias potential has converged. At this point, the bias, up to a multiplicative factor (which is specified by the tempering method) and an additive constant (which may be freely chosen) is the negative of the Landau free energy as a function of the CVs $\bm{\xi}$ in the simulated ensemble. The MetaD method, as originally proposed, continuously adds finite biases to the system throughout the entire duration of simulation, and thus detailed balance is not satisfied asymptotically,\citep{Barducci2008,Dama2014} though schemes similar to those applied in Wang-Landau sampling can be utilized which reduce the height of hills sequentially as specific milestones are reached.\citep{Wang2001,Wang2001a} A popular alternative scheme for mollifying the bias applied at late times, well-tempered metadynamics (WTMetaD), introduces a temperature enhancement $\Delta T$ such that the CV space is effectively sampled at temperature $T + \Delta T$ (where $T$ is the temperature of the system), and the heights $W(t)$ are based on the current bias applied at the system state:
\begin{equation}
   W(t) = W_0 e^{-V(\xi(t),t)/k_B\Delta T}
\end{equation}
Here, $W_0$ is a constant, and $\Delta T$ controls the rate of decay for the Gaussian height, and results in a converged bias which relates to the free energy as
\begin{equation}
    F(\vec{s}) = -\frac{T + \Delta T}{\Delta T} V_{\rm bias}(\vec{s})\;.
\end{equation}
\noindent where the multiplicative factor involving $\Delta T$ can be related to as the biasing factor $\Gamma$, 
\begin{equation}
    \Gamma = \frac{T + \Delta T}{T}
\end{equation}
via the equation\cite{Barducci2008}
\begin{equation}
	F(\vec{s}) = -\frac{\Gamma}{\Gamma-1}V_{\rm bias}(\vec{s})\;.
\end{equation}

\noindent The biasing factor $\Gamma$ is greater than one for $\Delta T > 0$, and implies a mollification, rather than exact cancellation, of the underlying free energy landscape by the bias potential. It governs the decay rate of the bias potential, if $\Gamma$ is infinitesimally small, the bias potential height decays to zero at no time and the sampling turns a classical unbiased simulation, on the other hand, if $\Gamma$ is infinitely large, the bias potential never decays, and the sampling becomes untempered metadynamics. Tempering methods are able to aid convergence when the relevant energy scales for sampling a landscape are roughly known {\it a priori} and comprehensive sampling of the relevant regions of CV space may be attained. This can be a problem in systems where a large dimensionality $d$ must be used for the CV space due to high free energy barriers or bottlenecks in the structural CVs of interest.

PBMetaD methods were proposed to alleviate this difficulty. PBMetaD draws inspiration from bias-exchange methods which bias on multiple CVs by switching configurations between multiple realizations of a molecular system undergoing MetaD using a parallel-tempering-like scheme.\citep{Piana2007a} These schemes facilitate exploration of new states while maintaining modest dimensionality in each individual free energy landscape. PBMetaD alters the algorithm in a novel way by exchanging the parallel copies of a molecular system for multiple biases probabilistically applied to the {\it same} system, so only a single simulation is necessary. With these modifications, an $n$-dimensional surface of interest can be exchanged for $n$ $1$-dimensional biases,\citep{Pfaendtner2015} which facilitates the convergence of each of the marginal biases. As exploration is facilitated by the weighted bias on each CV of interest, PBMetaD is also able to enhance exploration in each individual CV, and collectively drives a system toward new regions using the combined biases.

The expression for the the bias potential in a PBMetaD simulation is constructed through the following equation\citep{Prakash2018}:

\begin{equation}
   V(\bm{\xi},t)  = -k_BT\log(\sum\limits_\alpha e^{-\beta V_{PB}^\alpha (\xi_{\alpha},t)})\;,
\end{equation}

\noindent where

\begin{equation}
  V_{PB}^\alpha (\xi_{\alpha},t) = \sum\limits_{t'\leq t} W_0 e^{-\frac{V_{PB}^\alpha (\xi_{\alpha},t)}{k_B\Delta T}} e^{-\frac{[\xi_\alpha-s_{\alpha}(t')]^2}{2\sigma_{\alpha}^2}} P_{\alpha}(t')
\end{equation}

\noindent and

\begin{equation}
  P_{\alpha}(t') = \frac{e^{-\beta V_{PB}^\alpha (\xi_{\alpha},t')}}{\sum\limits_{\alpha}e^{-\beta V_{PB}^\alpha (\xi_{\alpha},t')}}
  \label{eqn:PBWeight}
\end{equation}

\noindent The equations are identical to WTMetaD up to the time-dependent parallel-bias weight factor $P_{\alpha}(t')$, which accounts for the conditional weight applied to each individual CV $\xi_{\alpha}$. This term accounts for the likelihood that a simulation at time $t$ occupies the parallel copy associated with the CV $\xi_{\alpha}$. Some rules have been developed for performing these simulations in a way which favors convergence of the bias within each variable which are inherited from MetaD and WTMetaD. The height $W$ and width $\sigma_\alpha$ should not be too low or too narrow and render sampling inefficient, nor too high or too wide that biases override important features in a rugged free energy landscape.

\begin{figure}[t!]
    \centering
    \includegraphics[width=0.45\textwidth]{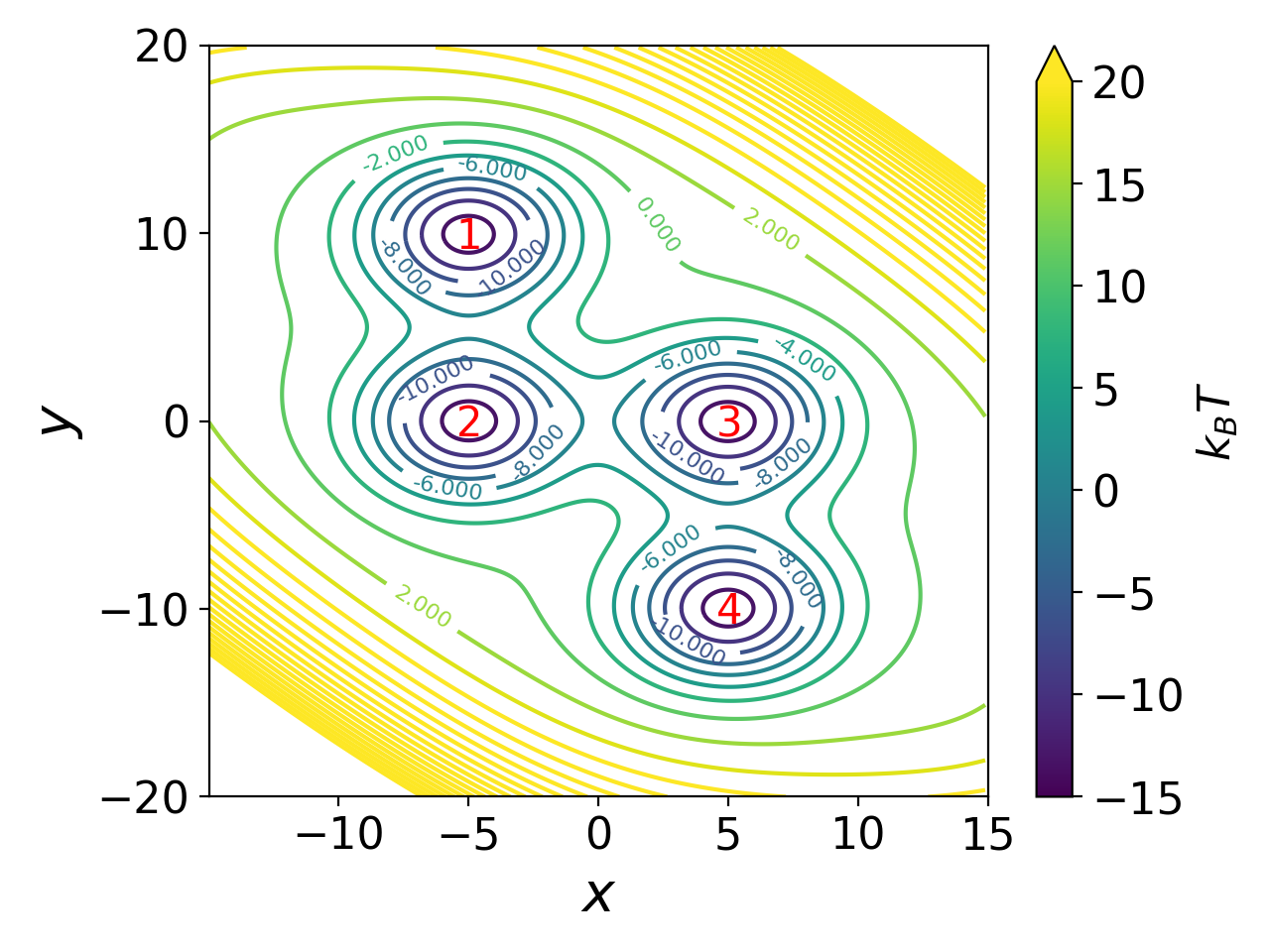}
    \caption{Four-Gaussian system potential surface. Under typical conditions, the system must proceed from the upper left basin to the lower right by hopping through the two basins at $y=0$. These intermediate states pose a problem for sampling when they are not explicitly delineated in the reaction coordinate of interest. Numerical labels in these basins are utilized for basin-correlation calculations.}
    \label{fig:4gauss}
\end{figure}

To test our hypothesis, we apply PBMetaD to two model system of $2$-dimensional energy surfaces, which are designed to have preferential ``hidden'' coordinates when examined from the standpoint of a simple structural collective variable such as the Cartesian $x$ or $y$ coordinates. Following Ref.~\citenum{Paz2018}, we begin with a four-well Gaussian potential surface. The system is plotted in Figure~\ref{fig:4gauss}. Such a surface models a system which exhibits two important functional conformations (at $y=-10$ and $y=10$), but must be activated at $y=0$ in order for a transition to happen. This model served as a pathological test case for conventional sampling techniques in Reference~\citep{Paz2018}. The explicit function used is

\begin{equation}
        \beta U(x,y)  = -15 \sum_{i=1}^4 \exp{\left(-{\frac{(x-x_i)^2+(y-y_i)^2}{15}}\right)}
        + \left(\frac{x+y}{15}\right)^8 + \left(\frac{x}{15}\right)^8 + \left(\frac{y}{15}\right)^8\;,
\label{eqn:4gauss}
\end{equation}

\noindent with Gaussian centers located at $(x_i,y_i) \in \left\{(-5,10), (-5,0), (5,0), (5,-10)\right\}$. Here, the quantity $\beta$ takes its typical value in statistical mechanics, $\beta = 1/k_{\rm B}T$

The shape of the potential presents several opportunities for optimizing sampling. In one sense, a one may use the parametric distance along a differentiable curve connecting all four basins defines an optimal reaction coordinate.\citep{Ribeiro2018,Sultan2018,E2005,Ren523} However, in a typical system this is not easily known. Hence, we sought a way to use simplified forms of coordinate combinations, some of which could be utilized to enhance sampling orthogonal to our chosen structural coordinate. We chose to explore sampling along a set of coordinates utilized in different linear combinations of $x$ and $y$. Gaussians are deposited every $100$ steps with a height $W = 0.1$ and width $\sigma= 0.3$ for every CV. In this case, $\Gamma$ is set to $10$, a value which was empirically determined to yield swift convergence in two-dimensional WTMetaD simulations of the surface. For each set of CVs, $100$ independent simulations are performed in order to compute average estimates for the convergence and exploration of each choice of CVs. Individual realizations run for $10^8$ timesteps with a timestep of $10^{-3}$ in reduced units. 

\begin{figure}[ht]
    \centering
    \includegraphics[width=0.45\textwidth]{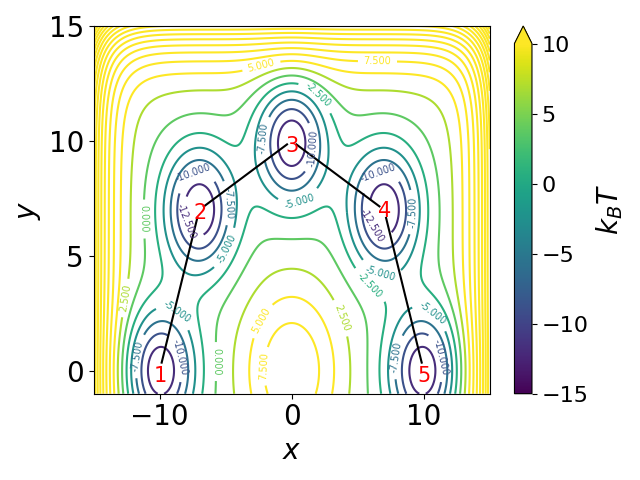}
    \caption{Arc model system potential surface, black straight lines are CVs in bias scheme involving segments. Numerical labels in these basins are utilized for basin-correlation calculations.}
    \label{fig:arcmodel}
\end{figure}

Additionally, a second model including ``hidden'' coordinates was also examined, which placed the centers of the Gaussian wells on an approximately semicircular arc. A barrier is placed at the origin as a disincentive to direct transfer between basins on the $x$-axis. The explicit potential was given by the equation

\begin{equation}
    \beta U(x,y)  = -15 \sum_{i=1}^5 \exp{\left(-{\frac{(x-x_i)^2+(y-y_i)^2}{7}}\right)}+ 10 \exp{\left(-{\frac{x^2+y^2}{15}}\right)}+ \left(\frac{x}{11}\right)^{10} + \left(\frac{y}{11}\right)^{10}\;,
    \label{eqn:arcmodel}
\end{equation}

\noindent where the Gaussian centers are located at  $(x_i,y_i) \in \left\{(-10,0), (-7,7), (0,10), (7,7), (10,7)\right\}$. Within this model, we regard the lower left basin as a reactant state and the lower right basin as a product state, with all other basins providing intermediate configurational states. This presents a useful alternative to the original four-basin surface, as the reaction coordinate in this case should be closely related to the polar angle, $\theta = \arccos(x/10)$, with a structural CV defined by $x$. We explore the influence of the additional coordinate, alongside a piecewise approximant informed by results gathered on the four-gaussian surface in Fig.~\ref{fig:4gauss}. The theoretical reaction coordinate should be well-approximated by the semicircle connecting all $5$ basins as this pathway costs the lowest energy. For this case, we explore the difference between parallel biasing which includes $\theta$ and biasing which includes a piecewise reaction coordinate created by biasing along the segments illustrated in Fig. \ref{fig:arcmodel}. 

In sampling this system, we again use Gaussians of height 0.1 $k_BT$, applied every $100$ steps. The widths are modified in this case; for linear CVs, we use $\sigma = 0.2$ for linear CVs ($x$ and all segments) and 0.02 for $\theta$. Analysis on this system is performed using $100$ parallel simulations, each of which runs for $2 \times 10^7$ timesteps with a timestep of $10^{-3}$ in reduced units. All simulations are carried out using LAMMPS\citep{Plimpton1995} with the PLUMED-2.5.3 plugin, an open-source, community-developed PLUMED library.\citep{Tribello2014,Bonomi2019}

\section{Results and discussion}

\begin{figure}[t]
    \centering
    \includegraphics[width=0.45\textwidth]{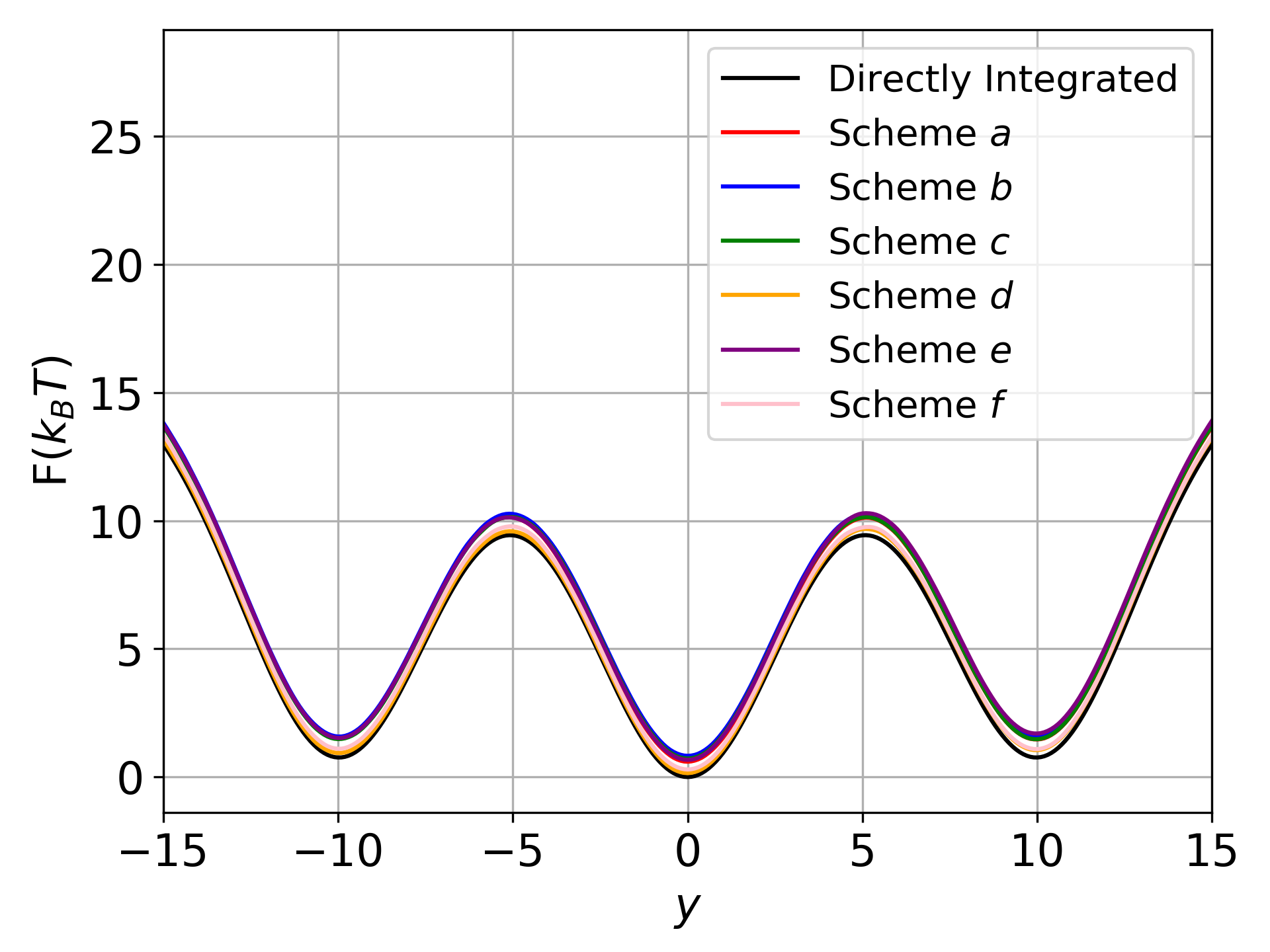}
    \caption{A comparison of the directly integrated free energy landscape of the four Gaussian potential with results from well-tempered MetaD and PBMetaD simulations having different combinations of collective variables as described in the text.}
    \label{fig:4gfel}
\end{figure}

For the model system illustrated by Fig.~\ref{fig:4gauss}, we choose $y$ as our primary structural CV as a reference to compare the effect of various bias schemes incorporating dynamical CVs including the orthogonal direction $x$ or linear combinations of $y$ and $x$. All simulations are either stock WTMetaD or PBMetaD methods on a collection of CVs which are linear combinations of the Cartesian coordinates. Collections with augmented CVs $s=y-x$ and $s=2y-x$ are of particular interest here, as they are able to separate all the relevant basins along the collective variable, and thus form simplified linear approximations of the true reaction coordinate. Importantly, as the exact potential energy surface is known in 2D, we can construct the free energy associated with projection of system configurations onto a collective variable through numerical integration of the partition function. This marginal free energy should be identical to the output of a converged MetaD simulation. Our results at late times for each of the CVs utilized appear to converge toward the expected marginal free energy, though there is some slight difference near the edges of the domain, as demonstrated in Figure~\ref{fig:4gfel}. Integrated free energy landscapes along each of the sampled CVs illustrating the marginal free energies in each system are included in the supplementary figures.

\begin{figure}[b!]
    \centering
    \includegraphics[width=0.45\textwidth]{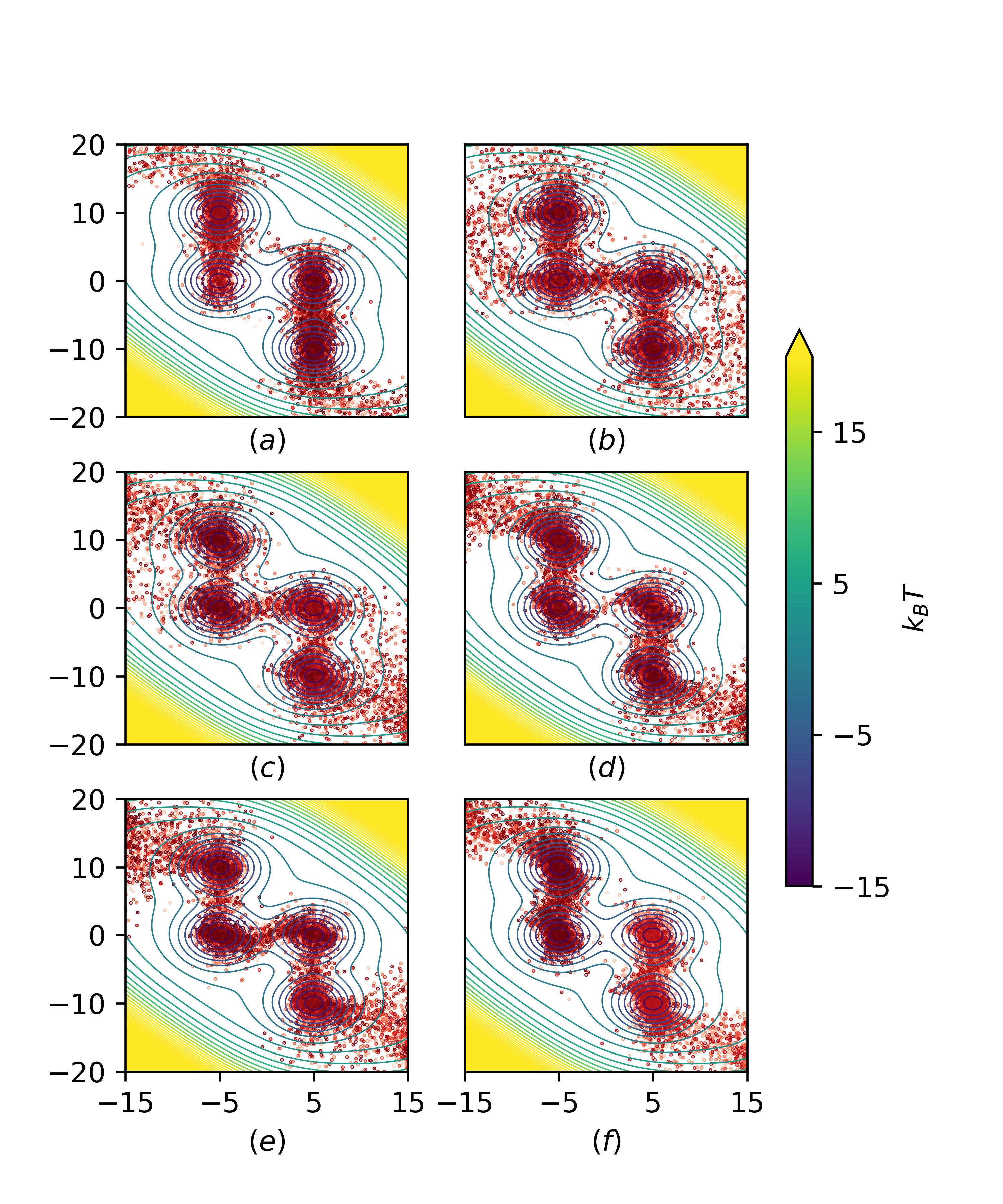}
    \caption{Typical trajectories on top of the 2D potential energy given in Eqn.~\ref{eqn:4gauss}. Each panel represents a WTMetaD or PBMetaD biasing scheme in Table.1 respectively. }
    \label{fig:4gtrj}
\end{figure}

Figure~\ref{fig:4gtrj} shows typical two-dimensional trajectories of the system on the model surface of Fig.~\ref{fig:4gauss}, in which the red dots represent locations where the system has visited, with darker shades of red used to illustrate later sampling times. In a classical MD simulation, the entire trajectory will lie in a single basin, as the energetic fluctuations necessary to escape from a $15k_BT$ deep minimum are rare. With a WTMetaD bias along only the $y$-direction, shown in Fig.~\ref{fig:4gtrj}(a), the trajectory demonstrates that the system is driven out of each basin in the $y$-direction, relying on contours of the free energy landscape and orthogonal fluctuations to sample the entire model surface. In particular, the transition between basin $2$ and basin $4$ does not go through basin 3, the minimum-energy pathway, instead climbing along $y$ to reach a higher energy level and slides downward into basin $4$. In this bias scheme, the system is not trapped in any state but does not sample all states in a dynamically optimal way: a critical transition pathway along $x$ connecting basin $2$ and $3$, which is hidden in the projection onto the $y$ coordinate, is not observed to contribute significantly to the sampled trajectory. With the introduction of an orthogonal coordinate $x$, trajectory changes significantly in Fig.~\ref{fig:4gtrj}(b): the cross-shaped trajectories in each basin combine to form a minimal energy pathway which captures all rare events. In Fig.~\ref{fig:4gtrj}(b), where the trajectory of a PBMetaD simulation biased along $x$ and $y$ is shown, the cross-shaped trajectory shows how the system escapes from each basin; importantly the introduction of an orthogonal factor of the structural CV $y$ enables the system to find the zigzag-shaped minimal energy pathway. This presents a question: how can choice of or augmentation by a different non-structural CV improve the overall sampling of the FEL? Since each CV comes with minimal overhead, a third CV can be added in PBMetaD without significant penalty. When the diagonal $y-x$ is added to the set used in Fig.~\ref{fig:4gtrj}(b), the primary effect is to alter the trajectories sampled within each basin; the transition pathways are still driven by sampling along the orthogonal variables, as shown in Fig.~\ref{fig:4gtrj}(c). 

Further biasing schemes are also explored, which demonstrate similar results. Figures~\ref{fig:4gtrj}(d)--(f) show trajectories using the collective variable sets $(y,y-2x)$, $(y,y-x)$, and $(y,2y-x)$, respectively, to explore how the relative angle between these CVs affects the overall sampling quality. The transition pathways are best explored when the set $(y,y-2x)$ in Fig.~\ref{fig:4gtrj}(d) is used. This is a bit surprising, as both this set and $(y,x)$ in Fig.~\ref{fig:4gtrj}(b) fail to separate the basins within the free energy landscape optimally, which was anticipated to be relevant in developing good dynamical coordinates. However, the orthogonal and near-orthogonal set of CVs offer a better piecewise approximation of the optimal reaction pathway in combination with the structural coordinate. While the optimal form of the CVs in facilitating transitions is surprising here, it is an extremely strong indication that our strategy of combining structural quantities of interest with quantities able to drive important dynamics should lead to highly successful PBMetaD sampling strategies.

\begin{table}
\begin{center}
    \begin{tabular}{ |c|c| }
    \hline 
     Scheme Label  & Bias Scheme \\
    \hline
    $a$ & $y$ \\
    \hline
    $b$ & $(y,x)$ \\
    \hline
    $c$ & $(y,x,y-x)$ \\
    \hline
    $d$ & $(y,y-x)$ \\
    \hline
    $e$ & $(y,y-2x)$ \\
    \hline
    $f$ & $(y,2y-x)$ \\
    \hline
    \end{tabular}
    \caption{Bias schemes used to sample the four Gaussian system. All simulations utilize either WTMetaD (one dimension) or PBMetaD (multiple dimensions).
    \label{tab:scheme}} 
\end{center}
\end{table}

\begin{figure}[ht]
    \centering
    \includegraphics[width=0.45\textwidth]{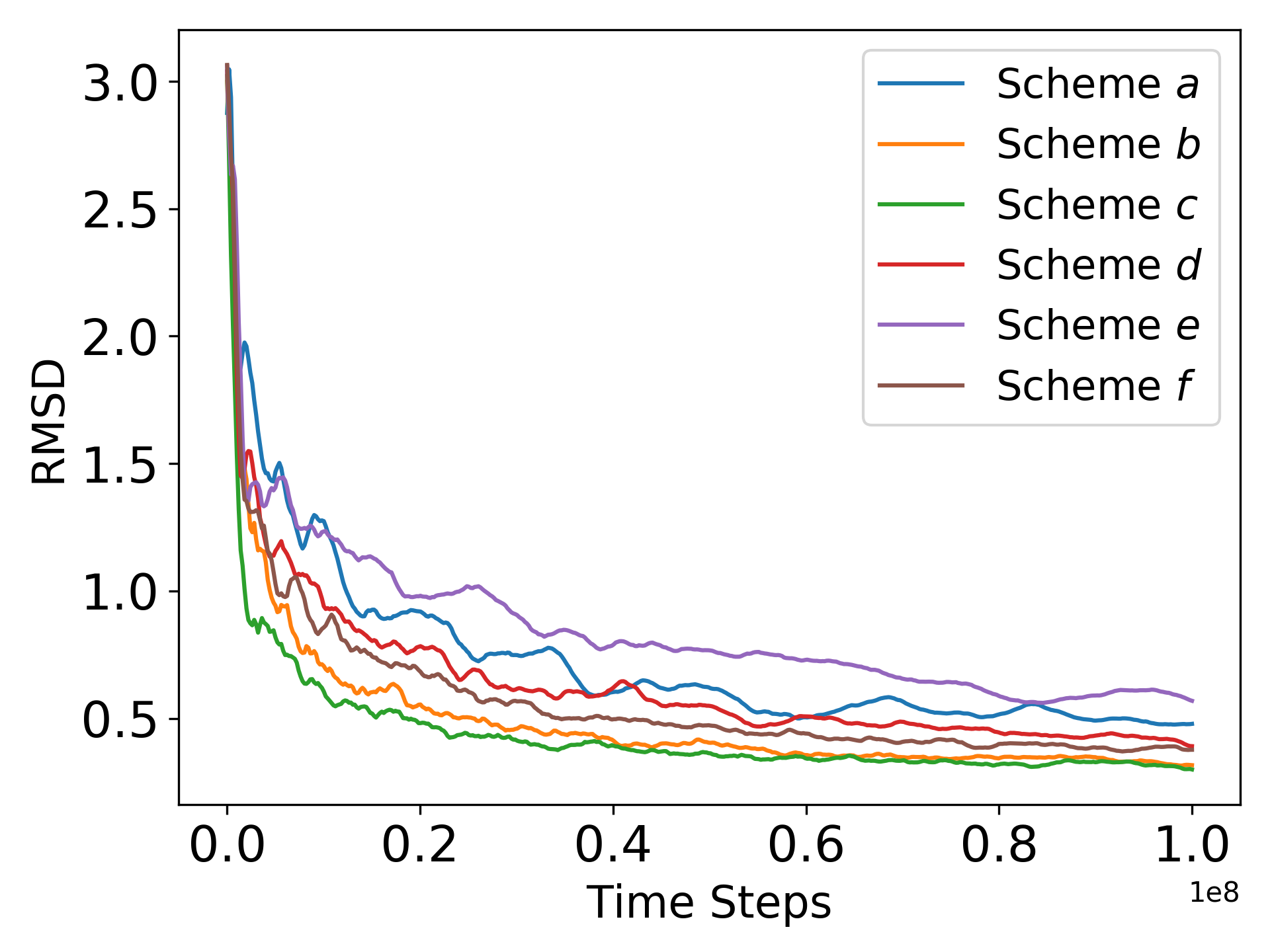}
    \caption{Time dependence of the root mean square deviation (RMSD) from the analytical free energy surface for different bias schemes applied to the four Gaussian surface. The region of interest is $(-15,15)$ along $y$ axis. See the main text for further discussion. Scheme numbers are given in Table~\ref{tab:scheme}.}
    \label{fig:4grmsd}
\end{figure}

\begin{figure}[ht]
    \centering
    \includegraphics[width=0.45\textwidth]{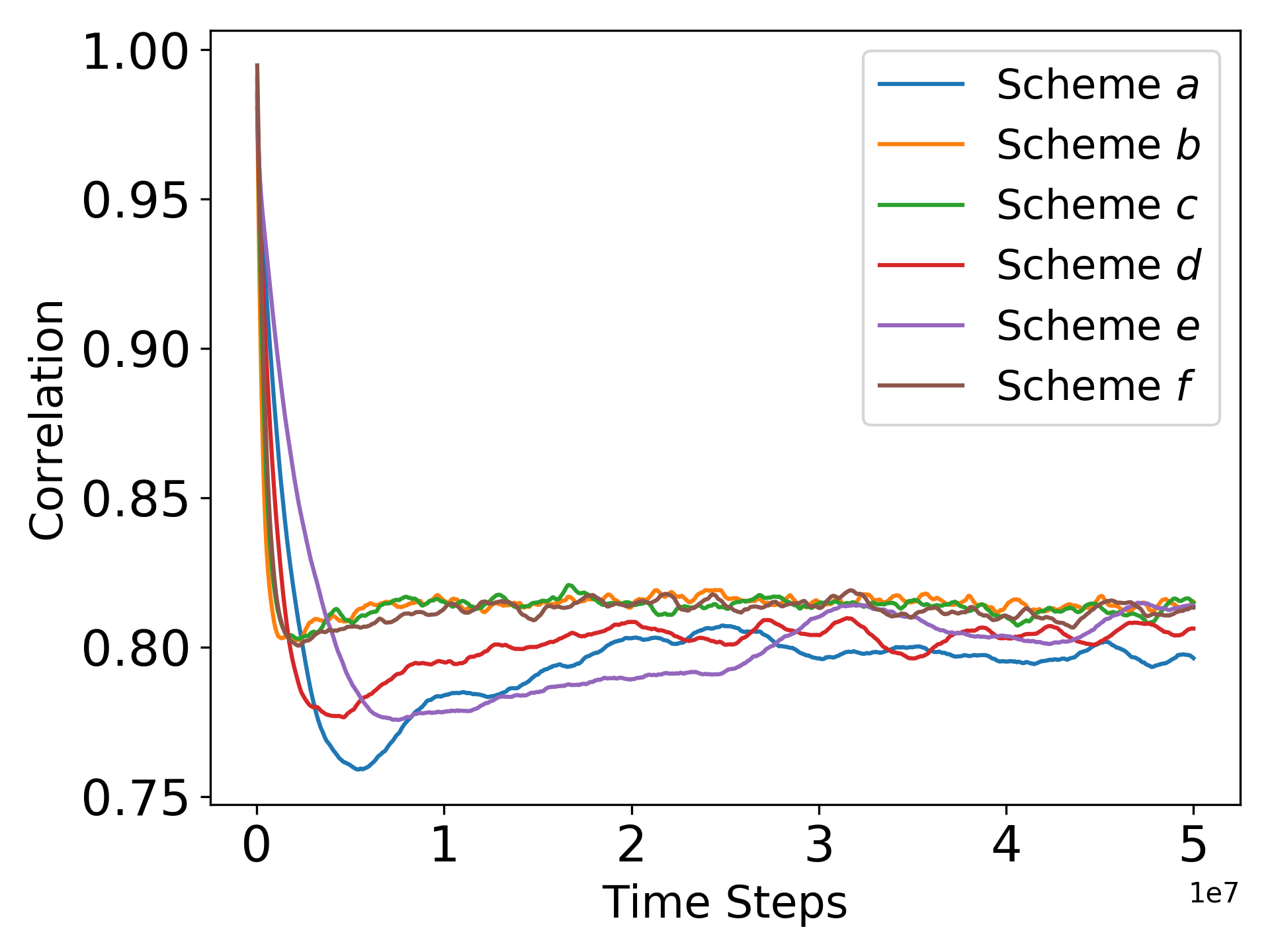}
    \caption{Time dependence of the basin correlation function proposed for the four Gaussian model. See the main text for further discussion. cheme numbers are given in Table~\ref{tab:scheme}.}
    \label{fig:4gcorr}
\end{figure}

Another factor in evaluating the effectiveness of sampling is convergence. Here we calculate the root-mean-square-deviation (RMSD) of reconstructed free energy landscapes along $y$ compared to the numerically integrated result from the model surface. The RMSD is defined as: 

\begin{equation}
    \mathrm{RMSD}   = \sqrt{\frac{1}{\Omega}\int_{\Omega}ds[(\widetilde{F}(s)-\langle\widetilde{F}(s)\rangle)-(F(s)-\langle F(s)\rangle)]^2},
    \label{eqn:RMSD}
\end{equation}

\noindent where $\Omega$ is the region of interest where free energy is integrated, $s$ denotes the CV, $\widetilde{F}(s)$ and $F(s)$ are directed integrated and reconstrcted free energies, respectively. While it should be noted that none of these RMSD calculations converge to zero error, convergence to the true free energy landscape is asymptotic~\cite{Dama2014}, and the timescales needed to ensure this difference is sufficiently small are difficult to obtain {\it a priori}. We can nevertheless obtain information about which scheme performs the best from the standpoint of convergence. For each bias scheme, the RMSD result is averaged over the ensemble of simulations performed with that biasing scheme, with the RMSD evaluated every $10^4$ timesteps. As shown in Fig.~\ref{fig:4grmsd}, bias schemes 2 and 3 approach zero the swiftest, indicating that samplings with an orthogonal CV $x$ have the best convergence. This makes sense, since these cases also do the best at driving transition between basins. However, it should be noted that all combinations of CVs do a better job than biasing only on $y$ alone. 

To further quantify the information extracted from trajectories, we study the basin index correlation of each bias scheme. The basin index correlation is defined as below.
\begin{equation*}
    C(\tau) = \frac{\langle i(t)\cdot i(t+\tau)\rangle}{\langle i(t)\cdot i(t)\rangle}
\end{equation*}
In which $i(t)$ is the basin index of the system at time $t$, which takes a value $i\in \left(1, 2, 3, 4\right)$ as defined in Fig.~\ref{fig:4gauss}. This offers an alternative measure of the efficacy of a sampling strategy, as it will approach a constant value at the point where a typical trajectory forgets its initial state. Methods which approach this value more quickly can thus be seen as more effective at driving transitions between the free energy basins present in the model. We observe in these cases that the correlation value eventually fluctuates around a value of about $0.8$. Our results, plotted in Fig.~\ref{fig:4gcorr} shows that the correlation of bias schemes ($x$,$y$), ($x$,$y$,$y-x$), and ($x$,$y-2x$) approach this steady value significantly faster than the rest. We thus conclude that sampling with CVs that piecewise approximates the reaction coordinate enables more comprehensive sampling, resulting in faster convergence toward the ideal free energy landscape, in further support of the prior results.

\begin{figure*}[t]
    \centering
    \includegraphics[width=0.90\textwidth]{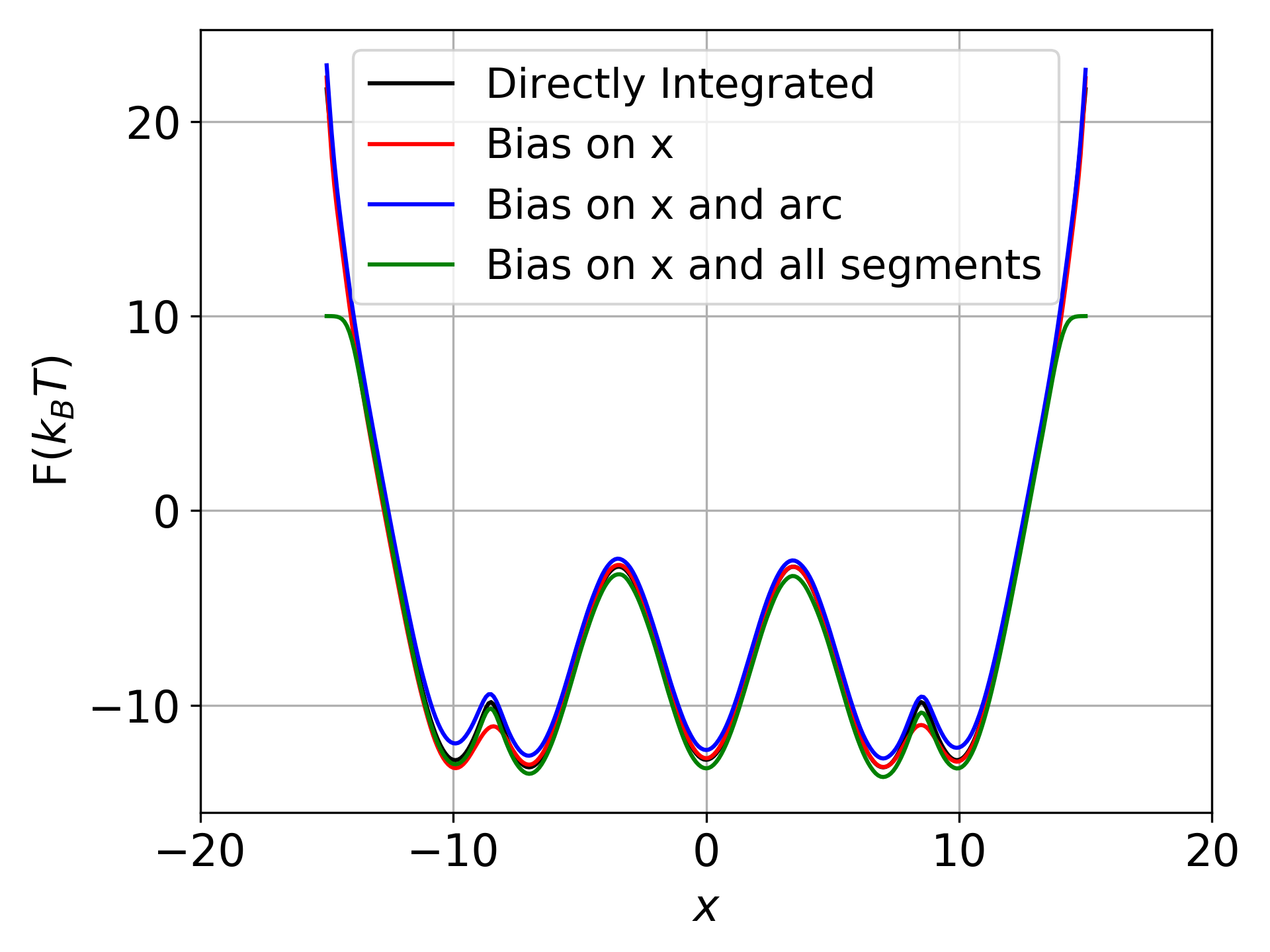}
    \caption{A comparison of the directly integrated free energy landscape of the arc model with results from well-tempered MetaD and PBMetaD simulations having different combinations of collective variables as described.}
    \label{fig:arcfe}
\end{figure*}

\begin{figure*}[t]
    \centering
    \includegraphics[width=0.90\textwidth]{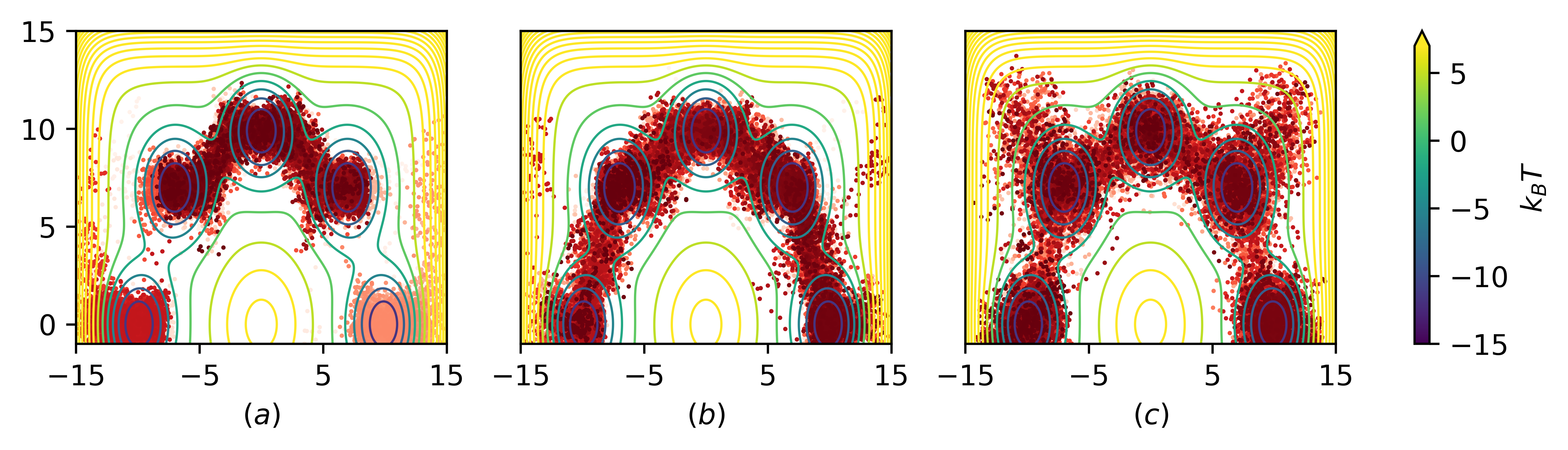}
    \caption{Two-dimensional trajectories of different bias schemes on the arc model surfaces. (a) bias on x only, (b) bias on x and $\theta$, (c) bias on x and all four segments connecting all basins. Strategies (b) and (c) lead to increased coverage of the relevant sampling regions and pathways}
    \label{fig:arctrj}
\end{figure*}

\begin{figure}[ht]
    \centering
    \includegraphics[width=0.45\textwidth]{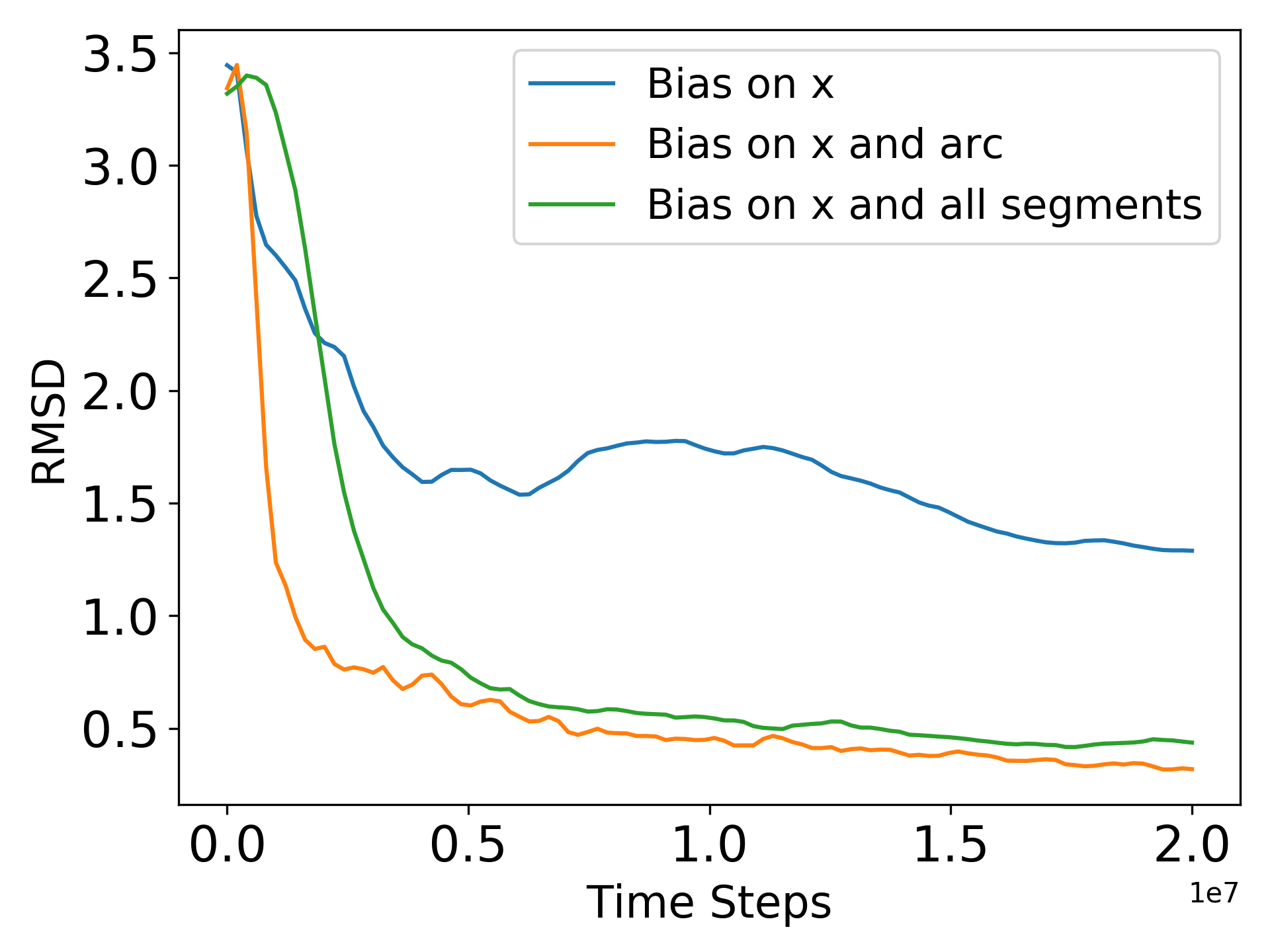}
    \caption{RMSD for different bias scheme applied to the arc model,  The region of interest is $(-13,13)$ along $x$ axis.}
    \label{fig:arcrmsd}
\end{figure}

\begin{figure}[ht]
    \centering
    \includegraphics[width=0.45\textwidth]{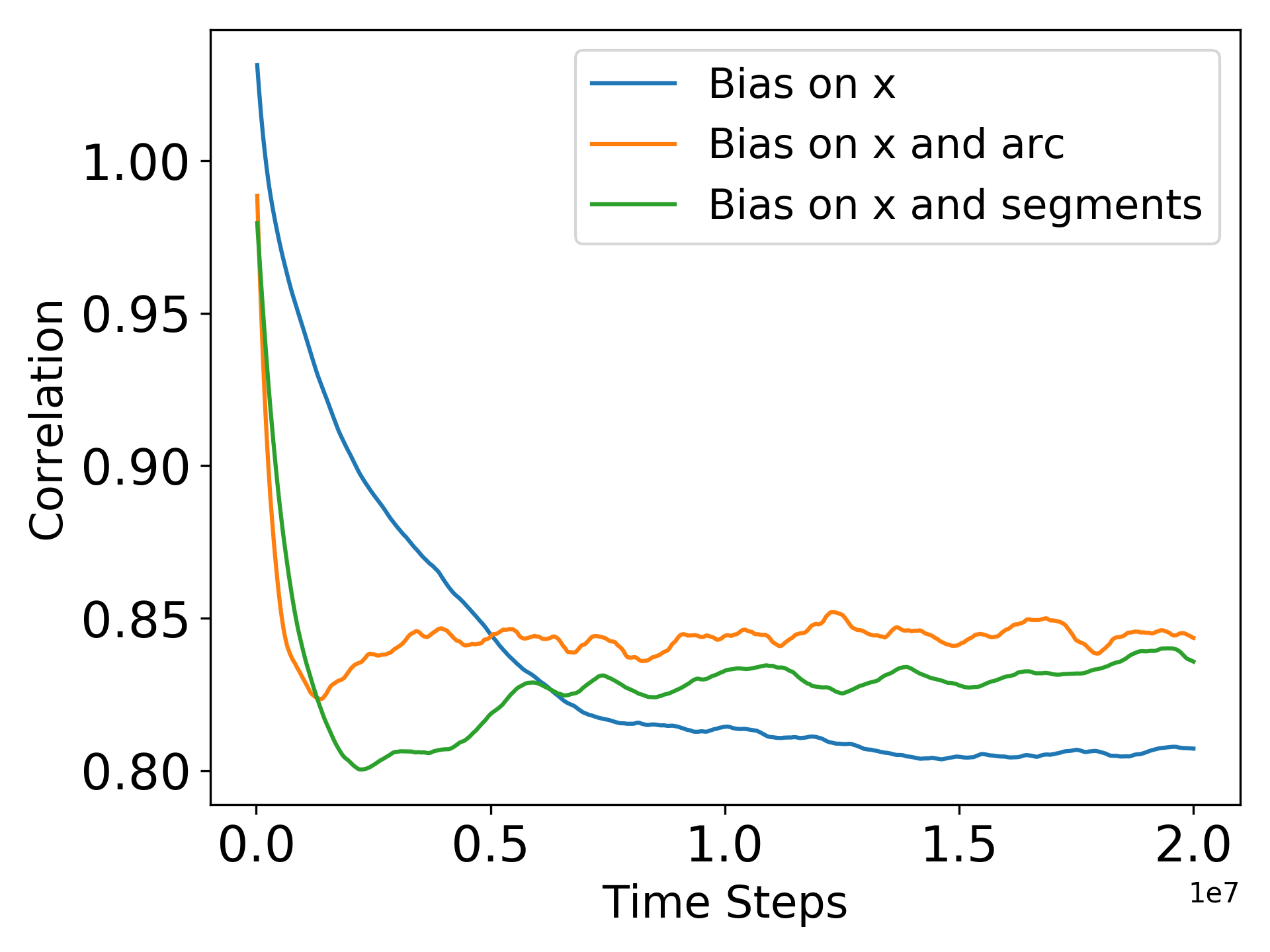}
    \caption{Basin correlation for different bias schemes applied to the arc model }
    \label{fig:arccorr}
\end{figure}

With this information in hand, we proceed to explore biasing in a situation where the reaction coordinate is not well aligned with the structural CV in any portion of the free energy landscape. For this we use the system depicted in Figure~\ref{fig:arcmodel}. We applied three bias schemes: $x$ only, $x$ with the semicircle arc, and $x$ with CVs describing the slope of the four segments connecting all five basins as illustrated in Fig.~\ref{fig:arcmodel}. Since the CV along the arc has units of radians rather than Cartesian dimension, we adjust the width of hills in this CV so that each deposited Gaussian covers approximately the same amount of the domain in this variable $\theta$. The reconstructed free energy profile is shown in Figure~\ref{fig:arcfe}, and integrated free energy landscapes along sampled dynamical CVs are included in the supplementary information.

We again explore typical trajectories for each biasing scheme in Fig.~\ref{fig:arctrj}. In Fig.~\ref{fig:arctrj}(a), where bias is applied only in the $x$ direction, the system inefficiently attempts to climb the hill at the origin or the bounding walls, and subsequently falls into basins at larger $y$-values along the arc. The tunnels between the reactant and product basins and each basin's nearest intermediate state, (which form an approximately $30^{\circ}$ angles with the $x$-axis) are essentially not utilized, limiting the biasing efficiency. In the other two cases, as shown in Fig.~\ref{fig:arctrj}(b) and Fig.~\ref{fig:arctrj}(c), transitions in the system are distributed along the semicircle, which is designed to be the dynamically favored pathway. However, when using the piecewise approximations in Figure~\ref{fig:arctrj}(c), some biasing is inefficiently used, as evidenced in the horn-like pattern traced out by the trajectory above the intermediate basins. 

Examination of these biasing schemes using our previously defined convergence and exploration criteria are given in Figs.~\ref{fig:arcrmsd} and~\ref{fig:arccorr}. It is clear from these results that while biasing with the true reaction coordinate performs the best, both in exploration and convergence measures, biasing which utilizes piecewise approximations of the surface is very close behind, despite the bias being spread among more CVs, thus reducing the overall weight applied to the structural CV at any point in time (see Eqn.~\ref{eqn:PBWeight}). This is highly suggestive that a strategy to obtain peak performance from the PBMetaD algorithm need not require full knowledge of the reaction coordinate(s) and rate limiting steps in a system. If we are aware of typical metastable states in between the relevant structural basins, choosing dynamical coordinates enabling exploration of the pathways between each individual basin, and thus a piecewise representation of the reaction coordinate partitioned among the aggregated CVs, offers behavior nearly as good as in the situation where the reaction coordinate is explicitly known. While the structure of the PBMetaD biases implies the weight applied to each is split among the different CVs, and thus encourages limiting the number of CVs utilized in a PBMetaD simulation somewhat, the evidence presented here clearly shows that exploration and convergence in MetaD simulations applied to a structural quantity of interest can be significantly enhanced through the use of a collection of CVs representing important underlying dynamics. Importantly, as may be observed in Figure~\ref{fig:arccorr}, exploration on $x$ alone leads to significantly slower decorrelation in the sampled states, which in general will lead to poor sampling and convergence of the free energy landscape.

\section{Conclusions} 

In this paper, we have explored the potential of parallel bias metadynamics
(PBMetaD) to drive sampling of the important structural degrees of freedom in
a system through inclusion of an ensemble of dynamical collective variables
which serve to drive a system over the transition states between metastable
configurations. We observe that our strategies of augmenting the structural
collective variable with a set of dynamical collective variables is able to
obtain a reconstructed free energy profile closely matching results from
direct integration. We further demonstrated how one may choose these dynamical
coordinates in a way which matches or closely matches the underlying reaction
coordinates of the system of interest to obtain the most efficient exploration
and convergence. This presents an alternative to strategies which aim to
accelerate MetaD and related sampling methods by obtaining the reaction
coordinate connecting structurally relevant basins through all important
transition states and metastable states---a highly nontrivial task. We
anticipate these results will be of great use to scientists studying complex
systems and transitions where some (but not all) dynamical information is
known, and combined with such as novel tools such as deep learning techniques
capable of obtaining reaction coordinates\citep{Ribeiro2018} and other slow
degrees of freedom\citep{Sidky2020} will allow a deeper understanding of the
thermodynamic properties, conformational states, and phase transitions
occurring in a broad range of complex systems.

\begin{acknowledgement}
  SH and JKW acknowledge support for this project from the United States National Science Foundation (Award No. DMR-1751988). MJQ and ECM were supported by the MICCoM Center at Argonne National Laboratory under a Computational Materials Science center grant from the Department of Energy, Basic Energy Sciences Division. 
\end{acknowledgement}

\bibliography{pb}

\newpage

\section{Appendix}
This section contains plots of the marginal free energy landscape along dynamical collective variables used to augment sampling. For more details see the main text.

\subsection{Marginal Free Energy Profiles in Four Gaussian Model}

\begin{figure}[h!]
    \centering
    \includegraphics[width=0.6\textwidth]{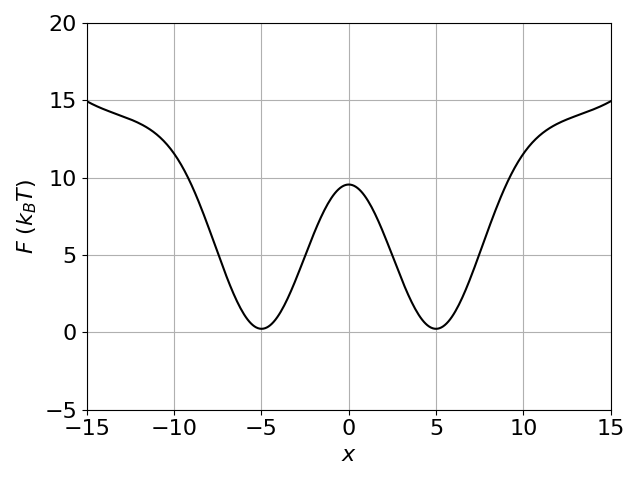}
    \caption{Marginal free energy along $x$ in Gaussian model}
    \label{fig:x}
\end{figure}

\newpage

\begin{figure}[h!]
    \centering
    \includegraphics[width=0.6\textwidth]{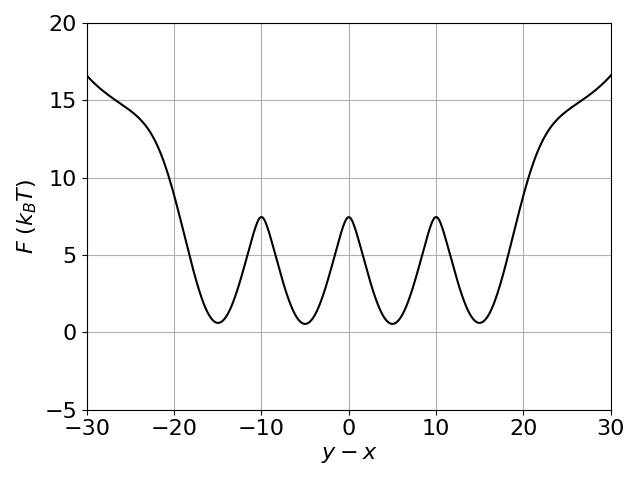}
    \caption{Marginal free energy along $y-x$ in Gaussian model}
    \label{fig:y-x}
\end{figure}

\newpage

\begin{figure}[h!]
    \centering
    \includegraphics[width=0.6\textwidth]{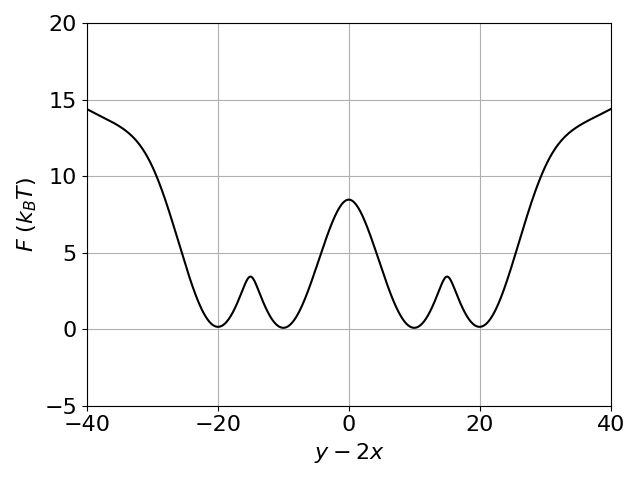}
    \caption{Marginal free energy along $y-2x$ in Gaussian model}
    \label{fig:y-2x}
\end{figure}

\newpage

\begin{figure}[h!]
    \centering
    \includegraphics[width=0.6\textwidth]{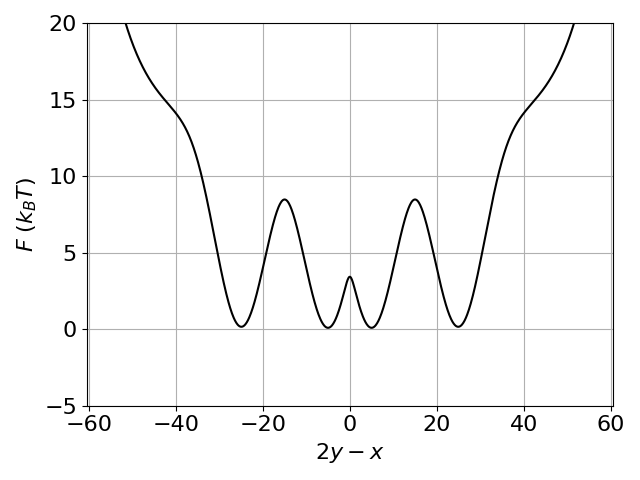}
    \caption{Marginal free energy along $2y-x$ in Gaussian model}
    \label{fig:2y-x}
\end{figure}

\newpage

\subsection{Marginal Free Energy Profiles in the Arc Model}

Due to symmetries, only three marginal surfaces are plotted. The lines connecting $3$ to $4$ and $4$ to $5$ may be obtained by symmetry from the lines connecting $2$ to $3$ and $1$ to $2$, respectively.

\begin{figure}[h!]
    \centering
    \includegraphics[width=0.6\textwidth]{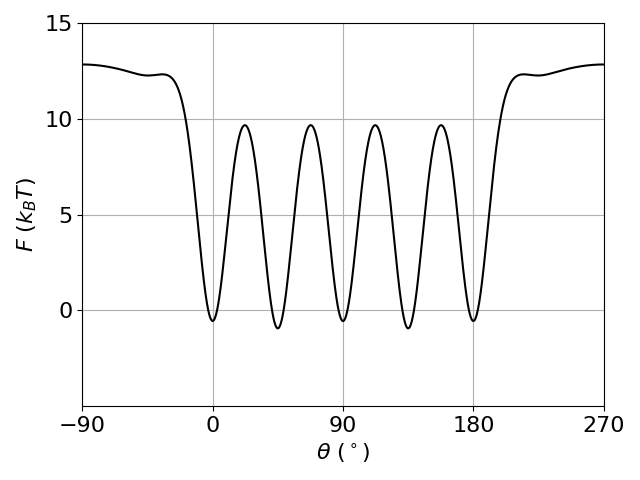}
    \caption{Marginal free energy along the arc in the arc model}
    \label{fig:arc}
\end{figure}

\newpage

\begin{figure}[h!]
    \centering
    \includegraphics[width=0.6\textwidth]{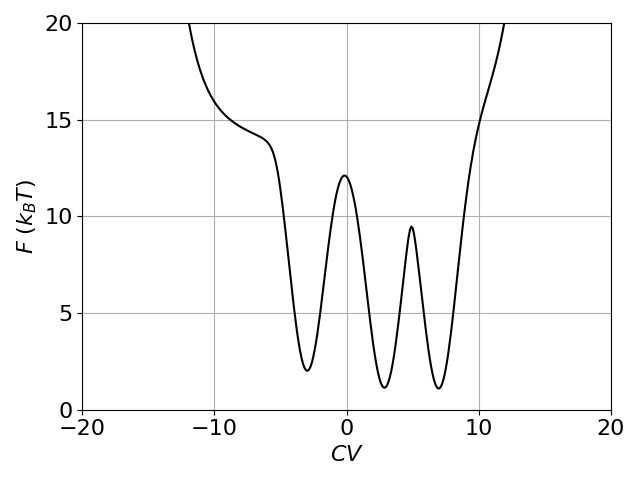}
    \caption{Marginal free energy along the dynamical CV connecting basin $1$ and $2$ in the arc model}
    \label{fig:seg12}
\end{figure}

\newpage

\begin{figure}[h!]
    \centering
    \includegraphics[width=0.6\textwidth]{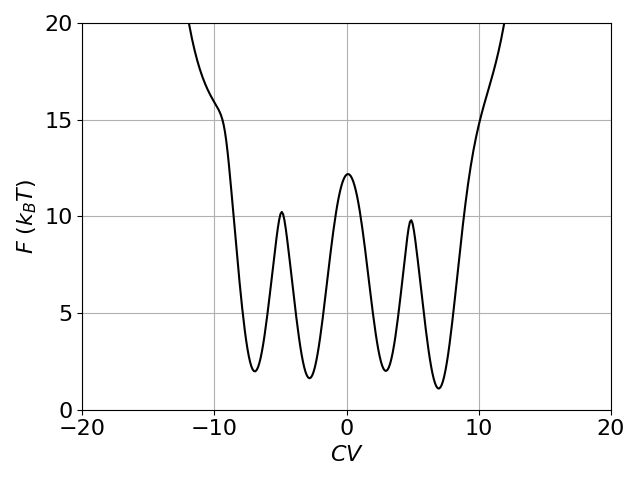}
    \caption{Marginal free energy along the dynamical CV connecting basin $2$ and $3$ in the arc model}
    \label{fig:seg23}
\end{figure}

\end{document}